\documentclass[a4paper,10pt,twoside]{cpc-hepnp}

\usepackage{multicol}
\usepackage{graphicx}
\usepackage{booktabs}
\usepackage{amssymb,bm,mathrsfs,bbm,amscd}
\usepackage[tbtags]{amsmath}
\usepackage{lastpage}

\def\undersim#1{\setbox9\hbox{${#1}$}{#1}\kern-\wd9\lower
    2.5pt \hbox{\lower\dp9\hbox to \wd9{\hss $_\sim$\hss}}}

\def\mv{{\mathbf v}}
\def\mr{{\mathbf r}}

\def\mr{{\mathbf r}}

\def\mk{{\mathbf k}}

\begin{document}
%\begin{CJK*}{GBK}{song}

\fancyhead[c]{\small Chinese Physics C~~~
Vol. 39, No. 3 (2015) 034103
}
%\fancyfoot[C]{\small xxxxxx-\thepage}

\footnotetext[0]{Received 26 June 2014}

\title{Relativistic effects on the back-to-back correlation functions of boson-antiboson pairs
in high energy heavy ion collisions\thanks{Supported by National Natural Science Foundation of
China (11275037)}}
\author{ZHANG Yong\quad YANG Jing
\quad ZHANG Wei-Ning$^{1)}$\email{wnzhang@dlut.edu.cn}}
\maketitle

\address{School of Physics and Optoelectronic Technology, Dalian University of Technology,
Dalian, Liaoning 116024, China\\}

\begin{abstract}
We calculate the back-to-back correlation (BBC) functions of relativistic boson-antiboson
pairs in high energy heavy ion collisions using the Monte Carlo method.  The relativistic 
effects on the BBC functions of $\phi\,\phi$ and $K^+K^-$ pairs are investigated.  
The investigations
indicate that the relativistic effects on the BBC functions of $K^+K^-$ pairs with large
momenta are significant, and the effect is sensitive to the particle freeze-out temperature.
\end{abstract}



%\footnotetext[0]{\hspace*{-3mm}\raisebox{0.3ex}{$\scriptstyle\copyright$}2014
%Chinese Physical Society and the Institute of High Energy Physics of the Chinese Academy
%of Sciences and the Institute of Modern Physics of the Chinese Academy of Sciences and IOP
%Publishing Ltd}%

\begin{multicols}{2}

\section{Introduction}
In the hot and dense hadronic sources created in high energy heavy ion collisions, the boson 
mass-shift due to the medium interactions might lead to a measurable back-to-back correlation
(BBC) of boson-antiboson pairs \cite{AsaCso96,AsaCsoGyu99}.  This medium-effective BBC is
different from the pure Bose-Einstein statistic correlations between the bosons with different
isospin \cite{AndWei91}, which are negligible as compared to the BBC in high energy heavy ion
collisions \cite{Gyu79,AsaCsoGyu99}.  Denote $a_\mk\, (a^\dagger_\mk)$ as the annihilation
(creation) operators of the freeze-out boson with momentum $\mk$ and mass $m$, and $b_\mk\,
(b^\dagger_\mk)$ as the annihilation (creation) operators of the corresponding quasiparticle
with momentum $\mk$ and shifted mass $m_{\!\!*}$ in the homogenous medium.  They are related
by the Bogoliubov transformation
\cite{AsaCso96,AsaCsoGyu99}
\begin{equation}
a_{\mk} = c_{\mk}\,b_{\mk} + s^*_{-\mk}\,b^\dagger_{-\mk},
\vspace*{-2mm}
\end{equation}
where
\vspace*{-2mm}
\begin{equation}
c_{\mk} = \cosh f_{\mk}\,, \hspace{4mm} s_{\mk} = \sinh f_{\mk}\,, \hspace*{4mm}
f_{\mk} = \frac{1}{2} \log (\omega_{\mk}/\Omega_{\mk}),
\end{equation}
\begin{equation}
\omega_\mk=\sqrt{\mk^2 + m^2}\,, \hspace*{0.5cm} \Omega_\mk=\sqrt{\mk^2 +m_{\!*}^2}\,.
\end{equation}
The BBC function is defined as \cite{AsaCso96,AsaCsoGyu99}
\begin{equation}
\label{BBCf}
C(\mk,-\mk) = 1 + \frac{|G_s(\mk,-\mk)|^2}{G_c(\mk,\mk) G_c(-\mk,-\mk) },
\end{equation}
where $G_c(\mk_1,\mk_2)$ and $G_s(\mk_1,\mk_2)$ are the chaotic and squeezed amplitudes,
respectively, as
\begin{equation}
G_c(\mk_1,\mk_2) = \sqrt{\omega_{\mk_1} \omega_{\mk_2}}\,\langle a^\dagger_{\mk_1}
a_{\mk_2} \rangle,
\end{equation}
\begin{equation}
G_s(\mk_1,\mk_2) = \sqrt{\omega_{\mk_1} \omega_{\mk_2} }\,\langle a_{\mk_1} a_{\mk_2}
\rangle,
\end{equation}
where $\langle \cdots \rangle$ means ensemble average.

In Refs. \cite{Padula06,Padula10,Padula10a}, S. Padula {\it et al.} studied the BBC
functions of $\phi\,\phi$ and $K^+K^-$ based on the non-relativistic formulism
\cite{Padula06} for local-equilibrium expanding sources in high energy heavy ion
collisions.  In this work we investigate the BBC functions of $\phi\,\phi$ and
$K^+K^-$ for the local-equilibrium expanding sources in a relativistic case, using
Monte Carlo calculations.  Our results indicate that the relativistic effects on
the BBC functions of $K^+K^-$ pairs with large momenta are significant, and the
effect is sensitive to the particle freeze-out temperature.

\section{BBC function for local-equilibrium expanding source}
\subsection{The formulas of BBC function}
For local-equilibrium expanding sources, $G_c(\mk_1,\mk_2)$ and $G_s(\mk_1,\mk_2)$
can be expressed as \cite{Sinyukov94,AsaCsoGyu99}
\begin{eqnarray}
&& \hspace*{-7mm} G_c({\mk_1},{\mk_2})\!=\!\int \frac{d^4\sigma_{\mu}(x)}{(2\pi)^3}
K^\mu_{1,2} e^{i\,q_{1,2}\cdot x}\,\! \Bigl\{|c'_{\mk'_1,\mk'_2}|^2\,n'_{\mk'_1,
\mk'_2}\nonumber \\
&& \hspace*{11mm} +\,|s'_{-\mk'_1,-\mk'_2}|^2\,[\,n'_{-\mk'_1,-\mk'_2}+1]\Bigr\},
\end{eqnarray}
\begin{eqnarray}
&& \hspace*{-7mm} G_s({\mk_1},{\mk_2})\!=\!\int \frac{d^4\sigma_{\mu}(x)}{(2\pi)^3}
K^\mu_{1,2}e^{2 i\,K_{1,2}\cdot x}\!\Bigl\{s'^*_{-\mk'_1,\mk'_2}c'_{\mk'_2,-\mk'_1}
\nonumber \\
&& \hspace*{3mm}\times n'_{-\mk'_1,\mk'_2}+\,c'_{\mk'_1,-\mk'_2}\,s'^*_{-\mk'_2,
\mk'_1}\,[\,n'_{\mk'_1,-\mk'_2} + 1] \Bigr\}.
\end{eqnarray}
Here, $d^4\sigma^{\mu}(x)=d^3\Sigma^{\mu}(x;\tau_f) F(\tau_f)d\tau_f$, $d^3\Sigma^{\mu}
(x;\tau_f)$ is the normal-oriented volume element depending on the freeze-out hypersurface
parameter $\tau_f$, $F(\tau_f)$ is the invariant distribution of the local time parameter,
$q^{\mu}_{1,2}=k^{\mu}_1-k^{\mu}_2$, $K^{\mu}_{1,2} =(k^{\mu}_1+k^{\mu}_2)/2$, and $\mk'_i$
is the local-frame momentum corresponding to $\mk_i~(i=1,2)$.  The other local variables
are:
\begin{equation}
c'_{\pm\mk'_1,\pm\mk'_2}=\cosh[\,f'_{\pm\mk'_1,\pm\mk'_2}\,],
\end{equation}
\begin{equation}
s'_{\pm\mk'_1,\pm\mk'_2}=\sinh[\,f'_{\pm\mk'_1,\pm\mk'_2}\,],
\end{equation}
\begin{eqnarray}
&&\hspace*{-5mm}f'_{\pm\mk'_1,\pm\mk'_2}=\frac{1}{2} \log \left[(\omega'_{\mk'_1}+
\omega'_{\mk'_2})/(\Omega'_{\mk'_1}+\Omega'_{\mk'_2})\right]\nonumber\\
&&\hspace*{8mm}=\frac{1}{2}\log\left[K^{\mu}_{1,2}u_{\mu}(x)/K^{*\nu}_{1,2}u_{\nu}(x)
\right]\nonumber\\
&&\hspace*{8mm}\equiv f_{\mk_1,\,\mk_2}(x),
\end{eqnarray}
\begin{eqnarray}
&&\hspace*{-7mm}\omega'_{\mk'_i}(x)=\sqrt{\mk'^2_i(x)+m^2}=k^{\mu}_i u_{\mu}(x)\nonumber\\
&&\hspace*{3mm}=\gamma_\mv\,[\,\omega_{\mk_i}-\mk_i\cdot\mv(x)\,],
\end{eqnarray}
\begin{eqnarray}
\label{Omp}
&&\hspace*{-7mm}\Omega'_{\mk'_i}(x)=\sqrt{\mk'^2_i(x)+m_*^2}\nonumber\\
&&\hspace*{3mm}=\sqrt{[k^{\mu}_i u_{\mu}(x)]^2-m^2+m_*^2}\nonumber\\
&&\hspace*{3mm}=k^{*\mu}_i u_{\mu}(x),
\end{eqnarray}
\begin{eqnarray}
\label{nkk}
&&\hspace*{-8mm}n'_{\pm\mk'_1,\pm\mk'_2}=\exp\left\{-\left[\frac{1}{2}\Big
(\Omega'_{\mk'_1}+\Omega'_{\mk'_2}\Big)-\mu_{1,2}(x)\right]\bigg/T(x)\right\}\nonumber\\
&&\hspace*{5.5mm}=\exp{\{-[K^{*\mu}_{1,2} u_\mu(x)-\mu_{1,2}(x)]/\,T(x)\}}\nonumber\\
&&\hspace*{5.5mm}\equiv n_{\mk_1,\,\mk_2}(x),
\end{eqnarray}
where, $K^{*\mu}_{1,2}=(k^{*\mu}_1+k^{*\mu}_2)/2$, $u^{\mu}(x)=\gamma_\mv[1,\mv(x)]$
is source velocity, $\mu_{1,2}(x)$ is pair chemical potential, and $T(x)$ is source
temperature, respectively.  Eq. (\ref{Omp}) gives the relationship between $k^{*\mu}
u_{\mu}(x)$ and $k^{\mu}u_{\mu} (x)$, which will be used in calculating $f_{\mk1,\mk2}
(x)$ and $n_{\mk1,\mk2}(x)$ for the expanding sources.

We take the source distribution in our calculations as
\begin{equation}
\label{rdis}
\rho(\mr)=C e^{-\mr^2/(2R^2)}\, \theta(r-2R),
\end{equation}
where $C$ is the normalization constant, and $R$ is the source radius.  The source velocity
is taken as
\begin{equation}
\mv(x)=\langle u \rangle \mr/(2R),
\end{equation}
where $\langle u \rangle$ is a velocity parameter \cite{Padula06}.
The emission time distribution is taken to be the typical exponential decay \cite{AsaCsoGyu99,Padula06,Padula10,Padula10a},
\begin{equation}
F(\tau) = \frac{\theta(\tau-\tau_0)}{\Delta t} \;e^{-(\tau-\tau_0)/\Delta t},
\end{equation}
where $\Delta t$ is a free parameter, and the effect of $F(\tau)$ on the BBC function
(\ref{BBCf}) is to multiply the second term by the factor $|{\widetilde F}(\omega_\mk
+ \omega_{-\mk}, \Delta t)|^2=[1+(\omega_\mk+\omega_{-\mk})^2 \Delta t^2]^{-1}$
\cite{AsaCsoGyu99,Padula06,Padula10,Padula10a}.  In the calculations of the BBC functions
of boson-antiboson pairs, we take $\mu_{1,2}=0$, and the parameters $R$ and $\Delta t$ are
taken to be 7 fm and 2 fm/$c$ \cite{Padula06,Padula10,Padula10a}.

For the considered source and with the sudden freeze-out assumption
\cite{AsaCsoGyu99,Padula06,Padula10,Padula10a}, we have
\begin{eqnarray}
\label{Gc}
&&\hspace*{-8mm}G_c(\mk_1,\mk_2)= \frac{K^0_{1,2}\,{\widetilde F}(\omega_{\mk 1}\!-
\omega_{\mk 2}, \Delta t)}{(2 \pi)^3 } \!\!\!\int \!\! d^3 r\,e^{i(\mk_1-\mk_2)\cdot
\mr}\nonumber\\
&&\hspace*{8mm}\times\, e^{-\mr^2\!/\!2R^2}\! \Big\{|c_{\mk1,\mk2}(x)|^2 n_{\mk1,\mk2}
(x)\nonumber\\
&&\hspace*{8mm}+\,|s_{\mk1,\mk2}(x)|^2 [n_{\mk1,\mk2}(x)+1] \Big\},
\end{eqnarray}
\begin{eqnarray}
\label{Gs}
&&\hspace*{-8mm}G_s(\mk_1,\mk_2)= \frac{K^0_{1,2}\,{\widetilde F}(\omega_{\mk 1}\!+
\omega_{\mk 2}, \Delta t)}{(2 \pi)^3 } \!\!\!\int \!\! d^3 r\,e^{i(\mk_1+\mk_2)\cdot
\mr}\nonumber\\
&&\hspace*{8mm}\times\, e^{-\mr^2\!/\!2R^2}\! \Big\{s^*_{\mk1,\mk2}(x)\,c_{\mk2,\mk1}
(x)\, n_{\mk1,\mk2}(x)\nonumber\\
&&\hspace*{8mm}+\,c_{\mk1,\mk2}(x)\, s^*_{\mk2,\mk1}(x)\, [n_{\mk1,\mk2}(x)+1] \Big\},
\end{eqnarray}
where
\begin{equation}
c_{\mk_1,\,\mk_2}(x)=\cosh[\,f_{\mk_1,\,\mk_2}(x)\,]=c'_{\pm\mk'_1,\pm\mk'_2},
\end{equation}
\begin{equation}
s_{\mk_1,\,\mk_2}(x)=\sinh[\,f_{\mk_1,\,\mk_2}(x)\,]=s'_{\pm\mk'_1,\pm\mk'_2},
\end{equation}
and
\begin{eqnarray}
\label{fbbc}
&&\hspace*{-5mm}C(\mk,-\mk)=1+\left(1+4\omega^2_\mk \Delta t^2\right)^{-1}
\Bigg|\!\int\!\!d^3r e^{\frac{-r^2}{2R^2}} \cr
&&\hspace*{10mm}\times\Big[s^*_{\mk,-\mk}(x)\,c_{-\mk,\mk}(x) n_{\mk,-\mk}(x)\cr
&&\hspace*{10mm}+c_{-\mk,\mk}(x)\, s^*_{\mk,-\mk}(x)(n_{-\mk,\mk}(x) +1)\Big]
\Bigg|^{\!2}\cr
&&\hspace*{10mm}{\Bigg /}\!\int \!\! d^3r e^{\frac{-r^2}{2R^2}}\Big[|c_{\mk,\mk}(x)|^2
n_{\mk,\mk}(x) \cr
&&\hspace*{10mm}+|s_{\mk,\mk}(x)|^2 (n_{\mk,\mk}(x) +1)\Big]\cr
&&\hspace*{10mm}{\Bigg /}\!\int \!\! d^3r e^{\frac{-r^2}{2R^2}}\Big[|c_{-\mk,-\mk}(x)|^2
n_{-\mk,-\mk}(x)\cr
&&\hspace*{10mm}+|s_{-\mk,-\mk}(x)|^2 (n_{-\mk,-\mk}(x) +1)\Big].
\end{eqnarray}

\subsection{Relativistic and non-relativistic results}
In this subsection we present the BBC functions of $\phi\,\phi$ and $K^+K^-$ calculated
by the Monte Carlo method.  In the calculations of the integrations in Eq. (\ref{fbbc}), 
for a given $\mk$ we generate the magnitude of $\mr$
according to Eq. (\ref{rdis}) and choose its angle with respect to $\mk$ in random.  Then,
we calculate the integrands and sum them in the $|\mk|$ bin.  For the non-relativistic case,
we use the approximation,
\begin{eqnarray}
\label{approx}
&&\hspace*{-5mm}k^{\mu}u_{\mu}=\gamma_{\mv} \big[\,\omega_{\mk} - \mk \cdot \mr \langle
u\rangle/(2R)\,\big] \nonumber\\
&&\hspace*{3mm}\approx \left(1+\frac{\mv^2}{2}\right)\left[m+\frac{\mk^2}{2m}- \mk \cdot
\mr \frac{\langle u\rangle}{2R}\right] \nonumber\\
&&\hspace*{3mm}\approx \left(1+\frac{\langle u\rangle^2}{8R^2}\mr^2\right)m +\left[
\frac{\mk^2}{2m}- \mk \cdot\mr \frac{\langle u\rangle}{2R}\right]\nonumber\\
&&\hspace*{3mm}=m+\frac{1}{2m}\left\{\mk-\mr\frac{\langle u\rangle}{2R} m\right\}^2.
\end{eqnarray}

We show in Fig. \ref{phi3d} the BBC functions of $\phi\,\phi$ in non-relativistic [panels
(a) and (c)] and relativistic [panels (b) and (d)] cases for the fixed freeze-out temperature
$T=140$ MeV \cite{Padula06} and the velocity parameter $\langle u\rangle=$ 0 and 0.5.
The non-relativistic results are approximately consistent with the results calculated
by the formulism in Ref. \cite{Padula06} (see Fig. 2(c) and 2(d) in \cite{Padula06}),
where slightly different non-relativistic approximations are used.  One can see from Fig.
\ref{phi3d} that the relativistic effect decreases the peaks of the BBC functions.

\end{multicols}
\begin{center}
\vspace*{-4mm}
\includegraphics[width=16cm]{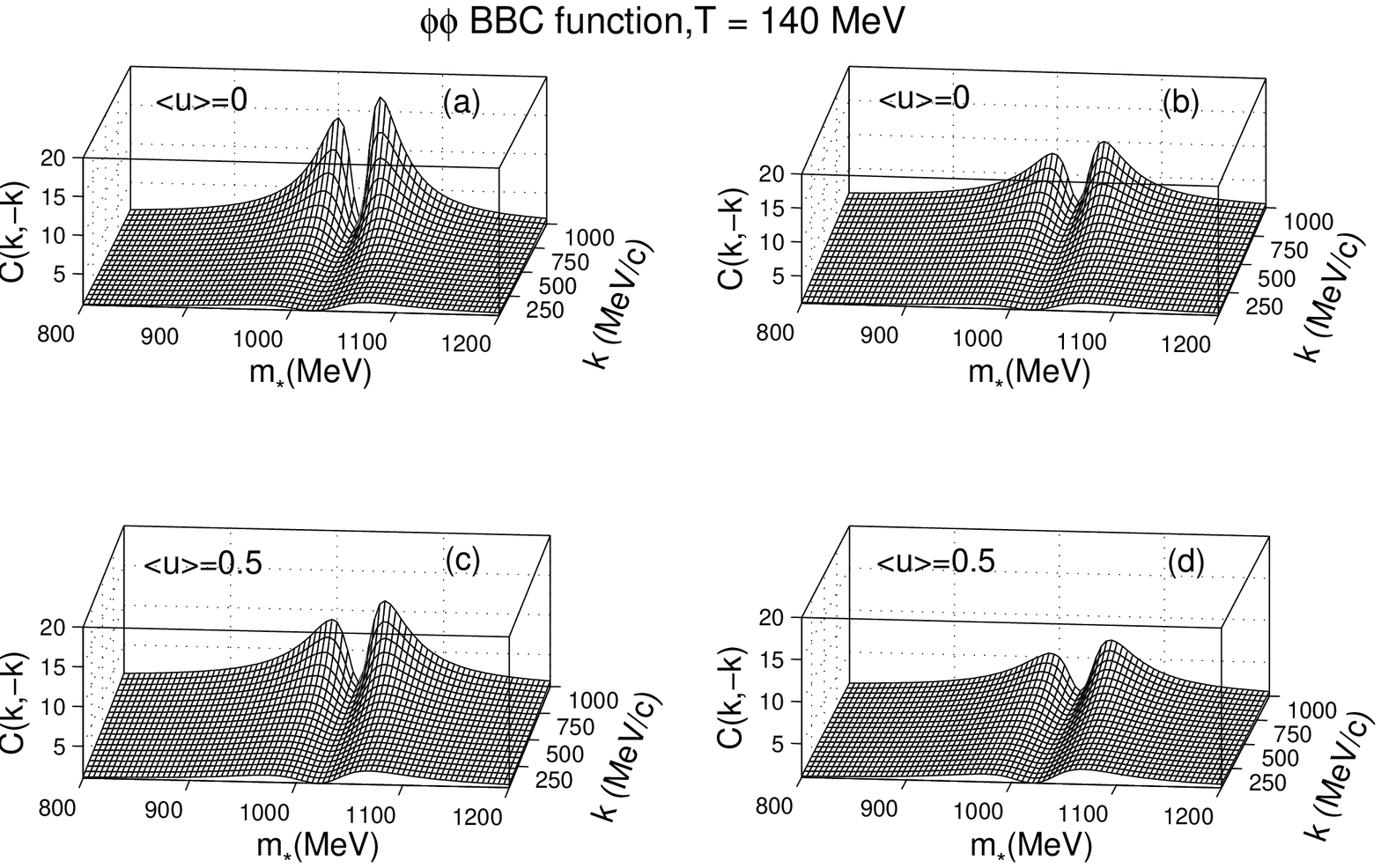}
\vspace*{0mm}
\figcaption{\label{phi3d} The BBC functions of $\phi\,\phi$ in the $m_*$-$k$ plane in
non-relativistic [(a) and (c)] and relativistic [(b) and (d)] cases for $T=140$ MeV,
$\langle u\rangle=$ 0 and 0.5. }
\end{center}

\begin{center}
\vspace*{3mm}
\includegraphics[width=10cm]{p2dT140.eps}
\vspace*{2mm}
\figcaption{\label{phi2d140} The BBC functions of $\phi\,\phi$ as a function of $m_*$ at
$k=$ 500 and 1000 MeV/$c$.  (a) and (c) are non-relativistic results.  (b) and (d) are
relativistic results. }
\end{center}
\begin{multicols}{2}

In Fig. \ref{phi2d140} we plot the BBC functions of $\phi\,\phi$ at $k=$ 500 and 1000
MeV$\!\!/c$ in non-relativistic [panels (a) and (c)] and relativistic [panels (b) and
(d)] cases.  At $k=500$ MeV$\!\!/\!c$, $\mk^2/ m^2_{\phi} \ll 1$, the differences
between the relativistic and non-relativistic BBC functions are very small.  However,
at the higher $k$, the differences between the relativistic and non-relativistic BBC
functions are larger.

\end{multicols}
\begin{center}
\vspace*{-3mm}
\includegraphics[width=16cm]{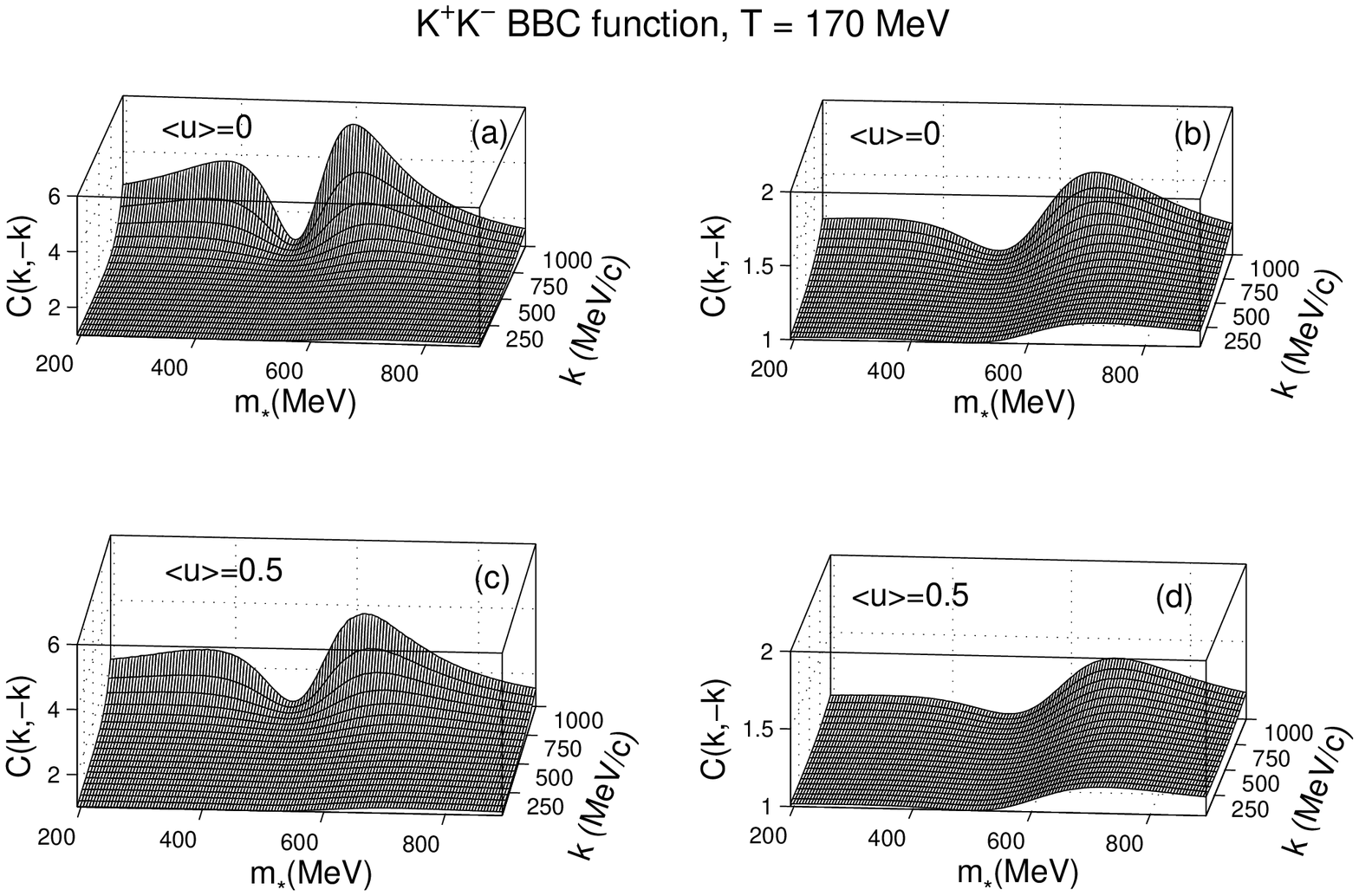}
\vspace*{0mm}
\figcaption{\label{k3d} The BBC functions of K$^+$K$^-$ in the $m_*$-$k$ plane in
non-relativistic [(a) and (c)] and relativistic [(b) and (d)] cases for $T=170$ MeV,
$\langle u\rangle=$ 0 and 0.5. }
\end{center}

\begin{center}
\vspace*{5mm}
\includegraphics[width=10cm]{k2dT170.eps}
\vspace*{2mm}
\figcaption{\label{k2d170} The BBC functions of K$^+$K$^-$ as a function of $m_*$ at
$k=$ 300 and 800 MeV/$c$.  (a) and (c) are non-relativistic results.  (b) and (d) are
relativistic results. }
\end{center}
\begin{multicols}{2}

In Fig. \ref{k3d} we show the BBC functions of K$^+$K$^-$ in non-relativistic [panels (a)
and (c)] and relativistic [panels (b) and (d)] cases for the fixed freeze-out temperature
$T=170$ MeV and the velocity parameter $\langle u\rangle=$ 0 and 0.5.  It can be seen that
the peaks of the relativistic BBC functions are suppressed.  Further, we plot in Fig.
\ref{k2d170} the BBC functions of K$^+$K$^-$ at $k=$ 300 and 800 MeV$\!\!/c$ in
non-relativistic [panels (a) and (c)] and relativistic [panels (b) and (d)] cases.  At the
smaller $k$, the relativistic effect on the BBC functions is small.  However, at the higher
$k$, the relativistic effect is significant. \\

\subsection{Relativistic effect on the BBC function}
To examine the relativistic effect on the BBC functions of boson-antiboson pairs, we define
$D_{\rm Rel}$ as the relative difference between the relativistic and non-relativistic BBC
functions $C^{\rm Rel}(\mk,-\mk)-C^{\rm Nrel}(\mk,-\mk)$ to $C^{\rm Rel}(\mk,-\mk)$,
\begin{eqnarray}
D_{\rm rel}=\frac{C^{\rm Rel}(\mk,-\mk)-C^{\rm Nrel}(\mk,-\mk)}{C^{\rm Rel}(\mk,-\mk)}.
\end{eqnarray}

\begin{center}
\vspace*{5mm}
\includegraphics[width=7cm]{zfp.eps}
\vspace*{2mm}
\figcaption{\label{Dp140} The relative difference $D_{\rm rel}$ as a function of $m_*$ for
the $\phi\,\phi$ BBC functions at $T=140$ MeV and with different momentum $k$ and $\langle
u\rangle$ values. }
\end{center}

In Fig. \ref{Dp140} we show the relative difference of $\phi\,\phi$ relativistic and
non-relativistic BBC functions at $T=140$ MeV.  For $\langle u\rangle =0$, the values
of $D_{\rm rel}$ are negative.  It means that the relativistic effect decreases the BBC
function.  The relativistic effect is small for low pair momentum and increases with pair
momentum.  At $k=0.8$ GeV/$c$, the relativistic effect may decrease the peak of BBC function
by 30\%.  For $\langle u\rangle =0.5$, the values of $D_{\rm rel}$ are larger than those
for $\langle u\rangle =0$.  From Eq. (\ref{approx}) one can see that the nonzero $\langle
u\rangle$ equivalently reduces the momentum for the large $\mk$, because Lorentz boost leads
to a bigger weight for the smaller angle between $\mk$ and $\mr$.  So the difference
between the relativistic and non-relativistic BBC functions decreases for the equivalent
small pair momentum.  However, the magnitude of the second term in the curly brace in Eq.
(\ref{approx}) may almost be the same or even larger than $\mk$ magnitude for large $\langle
u\rangle$ and small $k$.  This case leads to the positive $D_{\rm rel}$ results in Figs.
\ref{Dp140}(b) and \ref{Dp140}(c).

\begin{center}
\vspace*{2mm}
\includegraphics[width=7cm]{zfk.eps}
\vspace*{2mm}
\figcaption{\label{Dk170} The relative difference $D_{\rm rel}$ as a function of $m_*$ for
the $K^+K^-$ BBC functions at $T=170$ MeV and with different momentum $k$ and $\langle
u\rangle$ values. }
\end{center}

In Fig. \ref{Dk170} we show the results of $D_{\rm rel}$ for $K^+K^-$ relativistic and
non-relativistic BBC functions at $T=170$ MeV.  Because kaon mass is smaller than $\phi$
mass, the difference between the $K^+K^-$ relativistic and non-relativistic BBC functions
are larger.  It means the relativistic effect is important for $K^+K^-$ BBC function.
For $k=0.8$ GeV/$c$ and $\langle u\rangle =0$, one can see that the relative difference
even reaches to 45\%.  For the largest $\langle u\rangle$, only the $D_{\rm rel}$ results
for $k=0.3$ GeV/$c$ are a little greater than zero at large $m_*$.  It is because the
small kaon mass decreases the contribution of the second term in the curly brace in Eq.
(\ref{approx}).  In Fig. \ref{Dk160} we show the results of $D_{\rm rel}$ for the $K^+K^-$
BBC functions at $T=160$ MeV.  One can see that the differences between the relativistic
and non-relativistic BBC functions become larger at the smaller temperature.  The peak
altitude of the BBC function is mainly determined by the particle distribution $n_{\mk_1,\,
\mk_2}$ \cite{AsaCso96,AsaCsoGyu99}.  For fixed $k$ the difference between the $k^{\mu}u_{\mu}$
values in the relativistic and non-relativistic cases is enlarged in $n_{\mk_1,\,\mk_2}$
at a small temperature [see Eq. (\ref{nkk})].

\section{Summary and conclusion}
We calculate the BBC functions of relativistic boson-antiboson pairs in high energy heavy 
ion collisions using the Monte Carlo method.  The relativistic effects
on the BBC functions of $\phi\,\phi$ and $K^+K^-$ pairs are investigated for different
source velocities and pair momentum values.  We find that the relativistic effect on the
$\phi\,\phi$ BBC functions at low pair momentum is small.  However, at pair momentum $k=
0.8$ GeV/$c$ and with zero source velocity, the effect may decrease the peak of BBC function
by 30\%.  For a large source velocity, the non-relativistic BBC functions of $\phi\,\phi$
appears as a slight distortion of lessening at small $k$ values.
Because kaon mass is smaller the relativistic effect on $K^+K^-$ BBC function is more
important than that on $\phi\,\phi$ BBC function.  For $k=0.8$ GeV/$c$ the maximum of
the relative difference $D_{\rm rel}$ for the $K^+K^-$ BBC functions at $T=170$ MeV may
reach 45\%.  On the other hand, we find that the difference between the relativistic
and non-relativistic BBC functions is sensitive to the particle freeze-out temperature.
It becomes larger at a smaller freeze-out temperature.
Further investigations of the BBC for hydrodynamical sources \cite{EfaZha05,EfaZha12} 
will be of interest.

\begin{center}
\vspace*{0mm}
\includegraphics[width=7cm]{zfk160.eps}
\vspace*{2mm}
\figcaption{\label{Dk160} The relative difference $D_{\rm rel}$ as a function of $m_*$ for
the $K^+K^-$ BBC functions at $T=160$ MeV and with different momentum $k$ and $\langle
u\rangle$ values. }
\end{center}

%\acknowledgments{We thank  $\cdots$.}

\end{multicols}

\vspace{10mm}

%\begin{multicols}{2}

%\subsection*{Appendices A}
%\begin{small}

%\noindent{\bf Subtitle}

%\end{small}

%\end{multicols}

\vspace{-1mm}
\centerline{\rule{80mm}{0.1pt}}
\vspace{2mm}

\begin{multicols}{2}

\end{multicols}

\clearpage

%\end{CJK*}
\end{document}